\documentclass[aps,prl,twocolumn,groupedaddress,showpacs,nofootinbib,final,superscriptaddress,preprintnumbers,floatfix]{revtex4}
\usepackage{graphicx}
\usepackage{amssymb}
\usepackage{stmaryrd}

\begin{document}


\newcommand{\nbb}[0]{{0\nu}\beta\beta}
\newcommand{\fp}[0]{f_\pi}
\newcommand{\eq}[1]{eq.~(\ref{#1})}
\newcommand{\eqs}[2]{eqs.~(\ref{#1,#2})}
\newcommand{\Eq}[1]{Eq.~(\ref{#1})}
\newcommand{\Eqs}[2]{Eqs.~(\ref{#1},\ref{#2})}
\newcommand{\fig}[1]{fig.~\ref{#1}}
\newcommand{\figs}[2]{figs.~\ref{#1},\ref{#2}}
\newcommand{\Fig}[1]{Fig.~\ref{#1}}
\newcommand{\qbb}[0]{Q_{\beta\beta}}
\newcommand{\tl}[0]{\text{L}}
\newcommand{\tr}[0]{\text{R}}
\newcommand{\nc}[0]{N_{\text{c}}}
\newcommand{\mw}[0]{M_{\text{W}}}
\newcommand{\mz}[0]{M_{\text{Z}}}
\newcommand{\mr}[0]{M_{\text{R}}}
\newcommand{\md}[0]{m_{\text{D}}}
\newcommand{\mn}[0]{m_\nu}
\newcommand{\lh}[0]{\Lambda_{\text{H}}}
\newcommand{\aif}[0]{a^{ff^\prime}_{i;ll^\prime}}
\newcommand{\gs}[0]{\Gamma_{\text{S}}=1}
\newcommand{\gps}[0]{\Gamma_{\text{PS}}=\gamma^5}
\newcommand{\lr}[0]{\Lambda_{\text{R}}}
\newcommand{\wrt}[0]{W_{\text{R}}}
\newcommand{\wl}[0]{W_{\text{L}}}
\newcommand{\ls}[0]{\Lambda_{\text{S}}}
\newcommand{\gf}[0]{G_{\text{F}}}
\newcommand{\mm}[0]{M_{\text{M}}}
\newcommand{\sst}[1]{{\scriptscriptstyle #1}}
\newcommand{\beq}{\begin{equation}}
\newcommand{\eeq}{\end{equation}}
\newcommand{\beqa}{\begin{eqnarray}}
\newcommand{\eeqa}{\end{eqnarray}}
\newcommand{\dida}[1]{/ \!\!\! #1}
\renewcommand{\Im}{\mbox{\sl{Im}}}
\renewcommand{\Re}{\mbox{\sl{Re}}}
\def\simge{\hspace*{0.2em}\raisebox{0.5ex}{$>$}
     \hspace{-0.8em}\raisebox{-0.3em}{$\sim$}\hspace*{0.2em}}
\def\simle{\hspace*{0.2em}\raisebox{0.5ex}{$<$}
     \hspace{-0.8em}\raisebox{-0.3em}{$\sim$}\hspace*{0.2em}}
\def\dn{{d_n}}
\def\de{{d_e}}
\def\datom{{d_{\sst{A}}}}
\def\grhobar{{{\bar g}_\rho}}
\def\gpibar{{{\bar g}_\pi^{(I) \prime}}}
\def\gpibarz{{{\bar g}_\pi^{(0) \prime}}}
\def\gpibaro{{{\bar g}_\pi^{(1) \prime}}}
\def\gpibart{{{\bar g}_\pi^{(2) \prime}}}
\def\mx{{M_X}}
\def\mrho{{m_\rho}}
\def\qpv{{Q_{\sst{W}}}}
\def\lamtv{{\Lambda_{\sst{TVPC}}}}
\def\lamtvs{{\Lambda_{\sst{TVPC}}^2}}
\def\lamtvc{{\Lambda_{\sst{TVPC}}^3}}

\def\bra#1{{\langle#1\vert}}
\def\ket#1{{\vert#1\rangle}}
\def\coeff#1#2{{\scriptstyle{#1\over #2}}}
\def\undertext#1{{$\underline{\hbox{#1}}$}}
\def\hcal#1{{\hbox{\cal #1}}}
\def\sst#1{{\scriptscriptstyle #1}}
\def\eexp#1{{\hbox{e}^{#1}}}
\def\rbra#1{{\langle #1 \vert\!\vert}}
\def\rket#1{{\vert\!\vert #1\rangle}}

\def\lsim{{ <\atop\sim}}
\def\gsim{{ >\atop\sim}}
\def\nubar{{\bar\nu}}
\def\psibar{{\bar\psi}}
\def\Gmu{{G_\mu}}
\def\alr{{A_\sst{LR}}}
\def\wpv{{W^\sst{PV}}}
\def\evec{{\vec e}}
\def\notq{{\not\! q}}
\def\notl{{\not\! \ell}}
\def\notk{{\not\! k}}
\def\notp{{\not\! p}}
\def\notpp{{\not\! p'}}
\def\notder{{\not\! \partial}}
\def\notcder{{\not\!\! D}}
\def\notA{{\not\!\! A}}
\def\notv{{\not\!\! v}}
\def\Jem{{J_\mu^{em}}}
\def\Jana{{J_{\mu 5}^{anapole}}}
\def\nue{{\nu_e}}
\def\mns{{m^2_{\sst{N}}}}
\def\me{{m_e}}
\def\mes{{m^2_e}}
\def\mq{{m_q}}
\def\mqs{{m_q^2}}
\def\mw{{M_{\sst{W}}}}
\def\mz{{M_{\sst{Z}}}}
\def\mzs{{M^2_{\sst{Z}}}}
\def\ubar{{\bar u}}
\def\dbar{{\bar d}}
\def\sbar{{\bar s}}
\def\qbar{{\bar q}}
\def\sstw{{\sin^2\theta_{\sst{W}}}}
\def\gv{{g_{\sst{V}}}}
\def\ga{{g_{\sst{A}}}}
\def\pv{{\vec p}}
\def\pvs{{{\vec p}^{\>2}}}
\def\ppv{{{\vec p}^{\>\prime}}}
\def\ppvs{{{\vec p}^{\>\prime\>2}}}
\def\qv{{\vec q}}
\def\qvs{{{\vec q}^{\>2}}}
\def\xv{{\vec x}}
\def\xpv{{{\vec x}^{\>\prime}}}
\def\yv{{\vec y}}
\def\tauv{{\vec\tau}}
\def\sigv{{\vec\sigma}}

\def\sst#1{{\scriptscriptstyle #1}}
\def\gpnn{{g_{\sst{NN}\pi}}}
\def\grnn{{g_{\sst{NN}\rho}}}
\def\gnnm{{g_{\sst{NNM}}}}
\def\hnnm{{h_{\sst{NNM}}}}
\def\xivz{{\xi_\sst{V}^{(0)}}}
\def\xivt{{\xi_\sst{V}^{(3)}}}
\def\xive{{\xi_\sst{V}^{(8)}}}
\def\xiaz{{\xi_\sst{A}^{(0)}}}
\def\xiat{{\xi_\sst{A}^{(3)}}}
\def\xiae{{\xi_\sst{A}^{(8)}}}
\def\xivtez{{\xi_\sst{V}^{T=0}}}
\def\xivteo{{\xi_\sst{V}^{T=1}}}
\def\xiatez{{\xi_\sst{A}^{T=0}}}
\def\xiateo{{\xi_\sst{A}^{T=1}}}
\def\xiva{{\xi_\sst{V,A}}}
\def\rvz{{R_{\sst{V}}^{(0)}}}
\def\rvt{{R_{\sst{V}}^{(3)}}}
\def\rve{{R_{\sst{V}}^{(8)}}}
\def\raz{{R_{\sst{A}}^{(0)}}}
\def\rat{{R_{\sst{A}}^{(3)}}}
\def\rae{{R_{\sst{A}}^{(8)}}}
\def\rvtez{{R_{\sst{V}}^{T=0}}}
\def\rvteo{{R_{\sst{V}}^{T=1}}}
\def\ratez{{R_{\sst{A}}^{T=0}}}
\def\rateo{{R_{\sst{A}}^{T=1}}}
\def\mro{{m_\rho}}
\def\mks{{m_{\sst{K}}^2}}
\def\mpi{{m_\pi}}
\def\mpis{{m_\pi^2}}
\def\mom{{m_\omega}}
\def\mphi{{m_\phi}}
\def\Qhat{{\hat Q}}
\def\FOS{{F_1^{(s)}}}
\def\FTS{{F_2^{(s)}}}
\def\GAS{{G_{\sst{A}}^{(s)}}}
\def\GES{{G_{\sst{E}}^{(s)}}}
\def\GMS{{G_{\sst{M}}^{(s)}}}
\def\GATEZ{{G_{\sst{A}}^{\sst{T}=0}}}
\def\GATEO{{G_{\sst{A}}^{\sst{T}=1}}}
\def\mdax{{M_{\sst{A}}}}
\def\mustr{{\mu_s}}
\def\rsstr{{r^2_s}}
\def\rhostr{{\rho_s}}
\def\GEG{{G_{\sst{E}}^\gamma}}
\def\GEZ{{G_{\sst{E}}^\sst{Z}}}
\def\GMG{{G_{\sst{M}}^\gamma}}
\def\GMZ{{G_{\sst{M}}^\sst{Z}}}
\def\GEn{{G_{\sst{E}}^n}}
\def\GEp{{G_{\sst{E}}^p}}
\def\GMn{{G_{\sst{M}}^n}}
\def\GMp{{G_{\sst{M}}^p}}
\def\GAp{{G_{\sst{A}}^p}}
\def\GAn{{G_{\sst{A}}^n}}
\def\GA{{G_{\sst{A}}}}
\def\GETEZ{{G_{\sst{E}}^{\sst{T}=0}}}
\def\GETEO{{G_{\sst{E}}^{\sst{T}=1}}}
\def\GMTEZ{{G_{\sst{M}}^{\sst{T}=0}}}
\def\GMTEO{{G_{\sst{M}}^{\sst{T}=1}}}
\def\lamd{{\lambda_{\sst{D}}^\sst{V}}}
\def\lamn{{\lambda_n}}
\def\lams{{\lambda_{\sst{E}}^{(s)}}}
\def\bvz{{\beta_{\sst{V}}^0}}
\def\bvo{{\beta_{\sst{V}}^1}}
\def\Gdip{{G_{\sst{D}}^\sst{V}}}
\def\GdipA{{G_{\sst{D}}^\sst{A}}}
\def\fks{{F_{\sst{K}}^{(s)}}}
\def\FIS{{F_i^{(s)}}}
\def\fpi{{F_\pi}}
\def\fk{{F_{\sst{K}}}}
\def\RAp{{R_{\sst{A}}^p}}
\def\RAn{{R_{\sst{A}}^n}}
\def\RVp{{R_{\sst{V}}^p}}
\def\RVn{{R_{\sst{V}}^n}}
\def\rva{{R_{\sst{V,A}}}}
\def\xbb{{x_B}}
\def\mlq{{M_{\sst{LQ}}}}
\def\mlqs{{M_{\sst{LQ}}^2}}
\def\lscal{{\lambda_{\sst{S}}}}
\def\lvect{{\lambda_{\sst{V}}}}
\def\PR#1{{{\em   Phys. Rev.} {\bf #1} }}
\def\PRC#1{{{\em   Phys. Rev.} {\bf C#1} }}
\def\PRD#1{{{\em   Phys. Rev.} {\bf D#1} }}
\def\PRL#1{{{\em   Phys. Rev. Lett.} {\bf #1} }}
\def\NPA#1{{{\em   Nucl. Phys.} {\bf A#1} }}
\def\NPB#1{{{\em   Nucl. Phys.} {\bf B#1} }}
\def\AoP#1{{{\em   Ann. of Phys.} {\bf #1} }}
\def\PRp#1{{{\em   Phys. Reports} {\bf #1} }}
\def\PLB#1{{{\em   Phys. Lett.} {\bf B#1} }}
\def\ZPA#1{{{\em   Z. f\"ur Phys.} {\bf A#1} }}
\def\ZPC#1{{{\em   Z. f\"ur Phys.} {\bf C#1} }}
\def\etal{{{\em   et al.}}}
\def\delalr{{{delta\alr\over\alr}}}
\def\pbar{{\bar{p}}}
\def\lamchi{{\Lambda_\chi}}
\def\qw0{{Q_{\sst{W}}^0}}
\def\qwp{{Q_{\sst{W}}^P}}
\def\qwn{{Q_{\sst{W}}^N}}
\def\qwe{{Q_{\sst{W}}^e}}
\def\qem{{Q_{\sst{EM}}}}
\def\gae{{g_{\sst{A}}^e}}
\def\gve{{g_{\sst{V}}^e}}
\def\gvf{{g_{\sst{V}}^f}}
\def\gaf{{g_{\sst{A}}^f}}
\def\gvu{{g_{\sst{V}}^u}}
\def\gau{{g_{\sst{A}}^u}}
\def\gvd{{g_{\sst{V}}^d}}
\def\gad{{g_{\sst{A}}^d}}
\def\gvftil{{\tilde g_{\sst{V}}^f}}
\def\gaftil{{\tilde g_{\sst{A}}^f}}
\def\gvetil{{\tilde g_{\sst{V}}^e}}
\def\gaetil{{\tilde g_{\sst{A}}^e}}
\def\gvqtil{{\tilde g_{\sst{V}}^e}}
\def\gaqtil{{\tilde g_{\sst{A}}^e}}
\def\gvutil{{\tilde g_{\sst{V}}^e}}
\def\gautil{{\tilde g_{\sst{A}}^e}}
\def\gvdtil{{\tilde g_{\sst{V}}^e}}
\def\gadtil{{\tilde g_{\sst{A}}^e}}
\def\delp{{\delta_P}}
\def\delzp{{\delta_{00}}}
\def\deld{{\delta_\Delta}}
\def\dele{{\delta_e}}
\def\lnew{{{\cal L}_{\sst{NEW}}}}
\def\osffp{{{\cal O}_{7a}^{ff'}}}
\def\oszg{{{\cal O}_{7c}^{Z\gamma}}}
\def\osgg{{{\cal O}_{7b}^{g\gamma}}}


\def\slash#1{#1\!\!\!{/}}
\def\beq{\begin{eqnarray}}
\def\eeq{\end{eqnarray}}
\def\bea{\begin{eqnarray*}}
\def\eea{\end{eqnarray*}}
\def\NCA{\em Nuovo~Cimento}
\def\IJMP{\em Intl.~J.~Mod.~Phys.}
\def\NP{\em Nucl.~Phys.}
\def\PLB{{\em Phys.~Lett.}~B}
\def\JETPLett{{\em JETP Lett.}}
\def\PRL{\em Phys.~Rev.~Lett.}
\def\MPL{\em Mod.~Phys.~Lett.}
\def\PRD{{\em Phys.~Rev.}~D}
\def\PR{\em Phys.~Rev.}
\def\PRP{\em Phys.~Rep.}
\def\ZPC{{\em Z.~Phys.}~C}
\def\PTP{{\em Prog.~Theor.~Phys.}}
\def\Baryon{{\rm B}}
\def\Lepton{{\rm L}}
\def\sbar{\overline}
\def\stilde{\widetilde}
\def\st{\scriptstyle}
\def\sst{\scriptscriptstyle}
\def\vac{|0\rangle}
\def\argh{{{\rm arg}}}
\def\G{\stilde G}
\def\Wmess{W_{\rm mess}}
\def\NI{\stilde N_1}
\def\antivac{\langle 0|}
\def\infinity{\infty}
\def\mco{\multicolumn}
\def\epp{\epsilon^{\prime}}
\def\psibar{\overline\psi}
\def\nmess{N_5}
\def\chibar{\overline\chi}
\def\lagr{{\cal L}}
\def\drbar{\overline{\rm DR}}
\def\msbar{\overline{\rm MS}}
\def\conj{{{\rm c.c.}}}
\def\Et{{\slashchar{E}_T}}
\def\Etot{{\slashchar{E}}}
\def\mZ{m_Z}
\def\MPlanck{M_{\rm P}}
\def\mW{m_W}
\def\cbeta{c_{\beta}}
\def\sbeta{s_{\beta}}
\def\cW{c_{W}}
\def\sW{s_{W}}
\def\deltaeps{\delta}
\def\sigmabar{\overline\sigma}
\def\epsilonbar{\overline\epsilon}
\def\vep{\varepsilon}
\def\ra{\rightarrow}
\def\half{{1\over 2}}
\def\ko{K^0}
\def\be{\beq}
\def\ee{\eeq}
\def\bea{\begin{eqnarray}}
\def\eea{\end{eqnarray}}
\def\alr{A_{\sst{LR}}}

\def\centeron#1#2{{\setbox0=\hbox{#1}\setbox1=\hbox{#2}\ifdim
\wd1>\wd0\kern.5\wd1\kern-.5\wd0\fi
\copy0\kern-.5\wd0\kern-.5\wd1\copy1\ifdim\wd0>\wd1
\kern.5\wd0\kern-.5\wd1\fi}}
\def\ltap{\;\centeron{\raise.35ex\hbox{$<$}}{\lower.65ex\hbox{$\sim$}}\;}
\def\gtap{\;\centeron{\raise.35ex\hbox{$>$}}{\lower.65ex\hbox{$\sim$}}\;}
\def\gsim{\mathrel{\gtap}}
\def\lsim{\mathrel{\ltap}}
\def\slashchar#1{\setbox0=\hbox{$#1$}           
   \dimen0=\wd0                                 
   \setbox1=\hbox{/} \dimen1=\wd1               
   \ifdim\dimen0>\dimen1                        
      \rlap{\hbox to \dimen0{\hfil/\hfil}}      
      #1                                        
   \else                                        
      \rlap{\hbox to \dimen1{\hfil$#1$\hfil}}   
      /                                         
   \fi}                                        %

\setcounter{tocdepth}{2}







{
\title{Neutrino mass constraints on $\mu$-decay and $\pi^0\to
  \nu \bar{\nu}$} 

\author{Gary Pr{\'e}zeau}\affiliation{Jet Propulsion
  Laboratory/California Institute of  Technology, 4800 Oak Grove Dr,
  Pasadena, CA 91109, USA}
\author{Andriy Kurylov} 
\affiliation{Kellogg Radiation Laboratory, California Institute of
Technology, Pasadena, CA 91125, USA}




\begin{abstract}
In this letter, we show that upper-limits on neutrino mass
translate into upper-limits on the class of neutrino-matter interactions
that can generate loop corrections to the neutrino mass matrix.  We apply our 
results to $\mu$- and $\pi$-decays and derive 
model-independent limits on six of the ten parameters used to
parametrize contributions to $\mu$-decay that do not belong to the
standard model.  
 These upper-limits provide improved constraints on the five Michel
 parameters,
 $\rho,\xi^\prime,\xi^{\prime\prime},\alpha,\alpha^\prime$, that
 exceed PDG constraints by at least one order of magnitude.  For 
$\pi^0\to \nu \bar{\nu}$  we find for the
branching ratio: $B(\pi^0\to\nu \bar\nu) < 10^{-10}$.


\end{abstract}



\maketitle
}

With the discovery of neutrino oscillations a few years
ago~\cite{Fukuda:1998mi,Ahmad:2002jz,Eguchi:2002dm}, the 
neutrino mass matrix has become a subject of intensive experimental and 
theoretical research as it provides a unique window into physics
beyond the standard model (SM). Indeed, the combination of
WMAP~\cite{Spergel:2003cb},
2DFGRF~\cite{Colless:2001gk}, and
neutrino
oscillation data yield an upper limit of 0.23~eV for the mass of an active 
neutrino.  The Planck mission~\cite{planckmission}, to be
launched in 2007, may further improve this limit to
$\sim$0.04~eV~\cite{Hannestad:2002cn}.   With masses of active neutrinos  
at least six orders of magnitude smaller than those of all other SM
fermions, neutrino masses are presumably generated at an energy scale
which significantly exceeds the
electroweak scale. At low energies, manifestations of such new physics, 
including neutrino masses, are suppressed by inverse powers of this
heavy scale.  For example, in the see-saw mechanism, neutrino 
masses are inversely proportional to the heavy right-handed neutrino mass, 
which can range from a few TeVs to $10^{13}$~GeV depending on the model.

The study of non-SM  neutrino-matter interactions may also shed 
light on physics beyond the SM. However, since neutrino-matter cross sections 
are generally small, direct observation of these interactions is
experimentally challenging. Moreover, since the number of candidates for 
physics beyond the SM is large, determining the most viable particle 
physics scenario is non-trivial. In view of this situation, 
model-independent constraints on non-SM neutrino-matter interactions
in combination with the study of the neutrino mass matrix
should prove a valuable tool in the search for new physics.

In this letter we point out a general connection between the 
neutrino mass and 
scalar, pseudoscalar, and tensor (S,PS,T) neutrino-matter
interactions. In particular, we show that under
minimal assumptions these chirality-changing interactions generate
contributions to neutrino mass through loop effects.  We do not make
any assumption about the dynamical origin of the neutrino mass.
Instead, we perform a phenomenological analysis and require that
such contributions to the mass not exceed the physical neutrino mass.
This allows us to place stringent constraints on
chirality-changing neutrino-matter couplings.  Our general conclusions
are then applied to the SM-forbidden decay of $\pi^0$ into a neutrino
and an anti-neutrino with the same helicity
($\pi^0\to\nu\bar\nu$) and to $\mu$-decay. In the
former case we show that the cosmological neutrino mass upper-limit constrains the
branching ratio for $\pi^0\to\nu \bar\nu$ to be $\sim 10^{4}$
times smaller than the best current experimental
limit~\cite{Auerbach:2003fz,Atiya:1991eg}.  
For $\mu$-decay, we derive improved
constraints on five out eleven Michel parameters (MPs) that
exceed current experimental limits by at least one order of magnitude
\cite{Hagiwara:fs}.  We also point out that a non-zero
measurement by TWIST~\cite{twist} of the MPs $\delta$ and $\rho$
could be used to make a statement about the neutrino mass that should
be consistent with cosmological limits extracted from WMAP and
Planck in combination with galaxy redshift surveys (GRS).  Finally, we observe that the non-SM chirality-changing
interactions cannot account for the deviation from the SM value of
the weak mixing angle reported by the NuTeV
collaboration~\cite{Zeller:2001hh}.

\begin{figure}
\resizebox{8cm}{!}{
\includegraphics{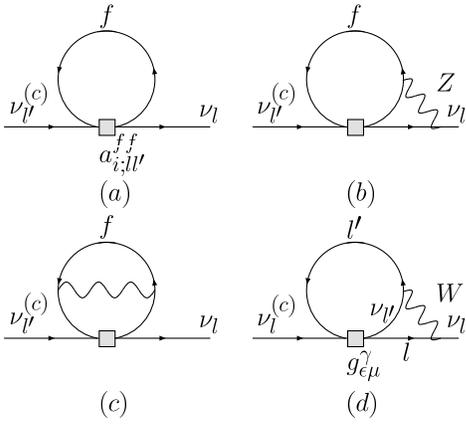}}
\vspace{-4.3cm}
\caption{One- and two-loop contributions to the neutrino mass (denoted
  $\delta m_\nu^{(1)},~\text{and}~\delta m_\nu^{(2)}$ respectively) 
generated by chirality-changing  neutrino-fermion interaction.  For
Majorana mass 
terms, $\nu_{l^\prime} \to \nu_{l^\prime}^c$ and there are two
additional diagrams similar to and of the same order as ($b$) and
($d$), where the weak bosons 
interact with $\nu_{l^\prime}^c$.} \label{fig:2loop}
\end{figure}

{\bf General Argument.}
The general chirality-changing effective neutrino-fermion interaction
can be written as
\beq \label{eq:interact} {\cal
  L}=\gf\sum_{l,l^\prime,f,f^\prime,i}a^{ff^\prime}_{i;l
  l^\prime}{\bar 
f}\Gamma_{i}f^\prime~{\bar \nu}_l \Gamma_{i}\nu_{l^\prime} +{\rm
  h.c.}~, \eeq 
where $i$=S,PS,T with $\Gamma_{\text{S}}=1$, $\Gamma_{\text{PS}}=
\gamma^5$, and $\Gamma_{\text{T}}=\sigma^{\mu\nu}$; the sum over 
$l,l^\prime$ runs over {\it active} neutrino flavors while the sum over
$f,f^\prime$ is over the SM 
charged fermions (this approach does not yield competitive
  constraints on neutrino-neutrino scattering), and
$a^{ff^\prime}_{i;l l^\prime}$ are
dimensionless constants parametrized in terms of the Fermi constant
$\gf=1.16637(1)\times 10^{-5}$~GeV$^{-2}$.
The chirality-changing
interaction in Eq.~(\ref{eq:interact}) generally contributes to
the neutrino mass via diagrams shown in Fig.~\ref{fig:2loop}.
Substituting $\bar{\nu}_l\Gamma_i\nu_{l^\prime}^c$ in \Eq{eq:interact} 
induces Majorana neutrino masses.

\Eq{eq:interact} is a general effective Lagrangian for neutrino-matter
interactions constructed from non-renormalizable operators (the
coupling constants have negative mass dimension as seen from the
overall factor of $\gf$).  Therefore, a new counterterm will be
needed for each operator to cancel divergences that may appear in
the evaluation of loop graphs.  The unique physical content of the loop
graphs resides in their non-analytical part. 
Analytical contributions can change with the
renormalization scheme used to make the graphs finite while
non-analytical terms remain the same.  Since we are only interested in
orders of magnitude, the only non-analytical contributions we will
consider are logarithms.

We evaluate leading logarithmic contributions to the
neutrino mass from the diagrams in Fig.~\ref{fig:2loop}.
The pseudoscalar and tensor neutrino-matter 
interactions are unconstrained to one-loop---the one-loop Feynman
diagrams with $\Gamma_i=\gamma^5,\sigma^{\mu\nu}$ give zero---while
the scalar interaction 
can be constrained by both the one- and two-loop contributions.  Since
we are interested in orders of
magnitude, we will not take into account the factors of ${\cal{O}}(1)$
associated with the different $\Gamma_i$'s.  The result is
\beqa \label{eq:delta-m} 
\delta m_\nu^{(1)} &\approx& \nc \gf a^{ff}_{\text{S};ll^\prime} {
m_f^3\over (4 \pi)^2}
\ln{\mu^2\over m_f^2}~,\nonumber \\
\delta m_\nu^{(2)}&\approx& 
g^2 \nc \gf a^{ff^\prime}_{i;ll^\prime} { 
(m_f~\text{or}~m_{f^\prime}) \mz^2\over (
4 \pi)^4}
\left( \ln{\mu^2\over \mz^2} \right)^2 ~, \eeqa
where the superscript in $\delta m_\nu^{(i)}$ indicates the
loop order, $\nc$ equals three for quarks and one for leptons,
$m_f$ is the mass of fermion $f$ (in \Eq{eq:delta-m}, it is the mass
of the 
fermion internal line that requires the chirality flip to couple to
the weak boson that is inserted), $g\cong 0.64$ is the SU(2$)_{\text{L}}$
coupling  
constant, $\mu$ is the renormalization scale, and where the subscripts
$ll^\prime$ are suppressed on the left-hand side of the matrix
equation~(\ref{eq:delta-m}).  Since $\delta
m_\nu^{(1)}/\delta m_\nu^{(2)}\sim 50 m_f^2/\mz^2$,  $\delta
m_\nu^{(1)}$ is 
negligible for all fermions except the top quark.
  Note that the loop expansion series converges since each loop
  order is suppressed by a numerical factor of 
$[\ln(\mu^2/\mz^2)/(4\pi)^2]^L$ where $L\ge 2$ is the loop order and the
  logarithm is of order ten as discussed below.
  Furthermore, the mass dependence of each loop diagram must be an
  expansion series in powers of $(m_f^2/\mz^2)^n m_f$ with 
$n=0,1,2,\dots$, and where the $n=0$ term appears only at second order
  with the exchange of a weak boson as in the diagrams of
  Fig.~\ref{fig:2loop}.  It follows that the $L=2$ diagrams with
  a weak boson are largest except for
  the case where $m_f=m_{\text{top}}$ as mentioned above.  
  We thus have the counterintuitive result that the 1-loop graph is generally subdominant.

The $(\ln\mu^2)^2$ factors in \Eq{eq:delta-m} appear because the diagrams are ultraviolet divergent; they are
compensated by the $\mu$-dependence of the neutrino mass counterterm
and the $\mu$-dependence of the
$a^{ff^\prime}_{i;ll^\prime}$'s deduced from the renormalization
group (RG) equations they satisfy.  Thus, in order to extract
constraints on the $a^{ff^\prime}_{i;ll^\prime}$'s, one must
choose a renomalization scale $\mu$.

This value of $\mu$
should exceed the mass of the heaviest particle included in the
effective field theory (EFT)---in our case $m_t$, the top quark mass---while at
the same time take into 
account the scale at which the onset of new physics might be expected.
We choose the renormalization scale to be
around 1~TeV, a scale often associated with physics beyond the
SM in many particle physics
models.  Since $\mu$ appears in a logarithm, our conclusions do not
  depend strongly on its precise value.  Note that the renormalization
  scale $\mu$ is far 
above the energy scale at which processes like $\mu$-decay and $\pi\to
  \nu\bar{\nu}$ occur.  In principle, the
  couplings appearing in \Eq{eq:delta-m} should be evolved down to
  $\sim$1~GeV using the appropriate  RG equations,
  but this can at most generate factors of ${\cal{O}}(1)$.  For
  example, the running of coupling constant associated with the
  four-quark operators in kaon decay, from the weak scale down to 
  $\mu\approx $1~GeV, generates only factors of 2~\cite{Donoghue:dd}.
  There is no reason to expect a more substantial change to the
  four-lepton or 
  quark-lepton operators of \Eq{eq:interact} when running $\mu$
  down to $\sim$100~MeV.  Thus, in the 
  model-independent analysis of this letter, there is no need to take
  the RG running of coupling constants into account.
  We emphasize that
  values of $\mu$ below the weak scale can not be substituted in
  \Eq{eq:delta-m}.  Below the weak scale, the dependence of the
  amplitude on $\mu$ becomes suppressed by inverse powers of the weak
  scale as required by the decoupling theorem.  See the section on QCD
  renormalization in Ref.~\cite{Hagiwara:fs} for a more detailed
  discussion of this point in the case of QCD.  Note that $\mu$ would
  not appear in a specific model where neutrino masses are calculated
  radiatively from finite diagrams.  In that case, the logarithms
  would instead have arguments of the form $M_\Lambda^2/\mz^2$ where
  $M_\Lambda$ would be the mass of a heavy particle in the model.

Below we use Eq.~(\ref{eq:delta-m}) to constrain $\aif$ by
requiring $\delta m_\nu\equiv\delta m_\nu^{(1)}+\delta m_\nu^{(2)}
\lesssim m_\nu$
where $m_\nu$ is the physical neutrino mass.  Since the graphs of
Fig.~\ref{fig:2loop} are divergent, there will be counterterms that
absorb the infinities.  In the absence of fine tuning and assuming
perturbation theory to be valid, the leading log 
contributions of the loop graphs should be no larger than the
physical value of neutrino masses.

We now apply our general results to non-SM $\pi^0$ and
$\mu$ decays. We adopt the upper limit of 0.7~eV on the sum of the 
neutrino masses from Ref.~\cite{Spergel:2003cb}, which translates
into the limit $m_\nu<0.23$~eV for individual neutrino masses when
neutrino oscillation constraints are included.

{\bf $\pi^0$-decay.}
We obtain from Eq.~(\ref{eq:delta-m})
$a^{qq}_{\text{PS};ll^\prime}< 10^{-3}$ for $q=u,d$. For the
calculation we Ä
used $m_f=m_u=m_d=(m_u+m_d)/2=4$~MeV~\cite{Hagiwara:fs} (constituent
quark masses are inappropriate when working at $\mu\sim 1$~TeV). We
can use this result to place an upper limit on the branching ratio
$B(\pi^0\to\nu\bar\nu)$. Starting from the neutrino-quark
interaction Lagrangian in Eq.~(\ref{eq:interact}) with
$\gps$ we obtain the effective interaction
\beq {\cal L}_{\pi\nu\bar\nu}=-\frac{\gf}{\sqrt{2}}{F_\pi m_\pi^2\over
m_u+m_d}\left(a^{uu}_{\text{PS};ll^\prime}-a^{dd}_{\text{PS};ll^\prime}\right)\pi^0\bar\nu_l
\gamma^5\nu_{l^\prime}~, 
\eeq 
where $F_\pi=92.4$~MeV is the pion decay constant and $m_\pi$ is the
pion mass. The above equation leads to the branching ratio
$B(\pi^0\to\nu\bar\nu)=
10^{-4}\left(a^{uu}_{\text{PS};ll^\prime}-
a^{dd}_{\text{PS};ll^\prime}\right)^2  <10^{-10}$, which is four
orders of magnitude stronger than the current best experimental
limit $B(\pi^0\to\nu\bar\nu)^{\rm Exp}<8.3\times 10^{-7}$
where $l=l^\prime=\mu$~\cite{Atiya:1991eg}
Our limit on $B(\pi^0\to\nu\bar\nu)$ improves by a further two orders
of magnitude if the possible Planck limit of $\mn<0.04$~eV is used
instead of $\mn<0.23$~eV.

{\bf $\mu$-decay.}
Muon decay can be described with the following effective
interaction
\cite{Hagiwara:fs} 
\beq
\label{eq:mu-decay}
{\cal L}_{\mu\to e\nu_\mu\bar\nu_e}={4\gf\over
\sqrt{2}}\sum_{{\gamma=S,V,T\atop \epsilon,\mu=R,L}}
g_{\epsilon\mu}^\gamma\bar
e_\epsilon\Gamma^\gamma\nu_e^n\bar\nu_{\text{muon}}^m\Gamma_\gamma \mu_\mu~,
\eeq
where $\gamma=S,V,T$ indicate, respectively, scalar, vector, and
tensor interactions and $\epsilon,\mu$=R,L indicate the chiralities of
the charged leptons. The chiralities $n$ and $m$ of the neutrinos are
determined by the values of $\gamma$, $\epsilon$, and $\mu$. The
constants $g_{\epsilon\mu}^\gamma$ parameterize the strength of
the corresponding phenomenological interactions and can be related to the
$a^{e\mu}_{i;\mu e}$ through Fierz transformations. In the SM,
$g_{LL}^V=1$ with the rest being zero.

Limits on
$g^{\text{S}}_{\text{RL}},~g^{\text{V}}_{\text{RL}},~g^{\text{T}}_{\text{RL}},~g^{\text{S}}_{\text{LR}},~g^{\text{V}}_{\text{LR}},~\text{and}~g^{\text{T}}_{\text{LR}}$
can be obtained from Fig.~\ref{fig:2loop}(d) and \Eq{eq:delta-m} with
$\mz\to \mw$ and $m_f=m_e$, the mass of the electron, for
$g^{\text{S}}_{\text{RL}},~g^{\text{V}}_{\text{RL}},~g^{\text{T}}_{\text{RL}}$ and
$m_f=m_\mu$, the mass of the muon, for
$g^{\text{S}}_{\text{LR}},~g^{\text{V}}_{\text{LR}},~\text{and}~g^{\text{T}}_{\text{LR}}$.  

\begin{table}\label{couplinglimits}
\begin{tabular}{|c|c|c|}
\hline
$g_{\epsilon\mu}^\gamma$ & Current Upper-limits & Upper-Limits from $m_\nu$
\\
\hline
$g^{\text{S}}_{\text{RL}}$ & 0.424 & $10^{-2}$
\\
\hline
$g^{\text{S}}_{\text{LR}}$ & 0.125 & $10^{-4}$
\\
\hline
$g^{\text{V}}_{\text{RL}}$ & 0.110 & $10^{-2}$
\\
\hline
$g^{\text{V}}_{\text{LR}}$ & 0.060 & $10^{-4}$
\\
\hline
$g^{\text{T}}_{\text{RL}}$ & 0.036 & $10^{-2}$
\\
\hline
$g^{\text{T}}_{\text{LR}}$ & 0.122 & $10^{-4}$
\\
\hline
\end{tabular}
\caption{Approximate upper-limits on the $g_{\epsilon\mu}^\gamma$'s from
  Ref.~\cite{Hagiwara:fs} in comparison to the ones derived from the
  loop-graph of Fig.~\ref{fig:2loop}(d) in combination with cosmological
  limits on the neutrino mass.} 
\end{table}

Our results are given in the third column of Tab.~I;
the second column displays current upper-limits from
Ref.~\cite{Hagiwara:fs}.  Except for $g^{\text{T}}_{\text{RL}}$, our
model-independent upper-limits are at least one order of magnitude
better than the ones appearing in Ref.~\cite{Hagiwara:fs}.
\begin{table}\label{Michel}
\begin{tabular}{|c|c|c|c|c|}
\hline
MP & PDG & WMAP/GRS  & TWIST/PSI  & Planck/GRS
\\
\hline
$\rho-3/4$ & 7  & 1 & 0.1  & 0.1
\\
\hline
$\eta$ & 33 & X & 0.1  &  X
\\
\hline
$\delta - 3/4$ & 7  & $10$ &  0.1 & 0.1
\\
\hline
$1-\xi\delta/\rho$ & $3.2^\dagger$  &  X  & 0.1  &  X
\\
\hline
$1-\xi^\prime$ & 80 & [$10$] & X  & [$4$]
\\
\hline
$1-\xi^{\prime\prime}$ & 58  & $10$ & X  & 0.1
\\
\hline
$\alpha/A$ & 9 & 0.001  & X  & 0.0001
\\
\hline
$\alpha^\prime/A$ & 8.8  & 0.001 & X  & 0.0001
\\
\hline
$a/A$ & $15.9^\dagger$ & 1 & X  & 0.1
\\
\hline
$a^\prime/A$ & 13 & 1 & X  & 0.1
\\
\hline
$c/A$ & $6.4^\dagger$  & 0.1 & X  &  0.01
\\
\hline
$c^\prime/A$ & 7.5 & 0.1   & X  & 0.01
\\
\hline
\end{tabular}
\caption{Order of magnitude upper-limits on the MPs.  All numbers should be
  multiplied by $10^{-3}$.  Note that $a,a^\prime,c,c^\prime$
  are not technically MPs, and instead belong to a set of parameters
  defined by Kinoshita and Sirlin~\cite{kinoshita57}.  PDG numbers are given
  in the second column at 95\% confidence level (CL) and
  90\%~CL (numbers with daggers).  The third column shows
  the order of magnitude limits 
  extracted from the  $g_{\epsilon\mu}^\gamma$'s given in Tab.~I and
  Ref.~\cite{Hagiwara:fs}.  The fourth column gives
  expected   order of magnitude 
  limits from the TWIST and PSI experiments~\cite{twist,PSI}.   The
  fifth column refers to improved limits on the MPs due to
  the anticipated data from the Planck Mission
  expected to constrain the upper-limit on the neutrino mass to around
  $\mn\lesssim 0.04$~\cite{Hannestad:2002cn}; see also
  Refs.~\cite{Lesgourgues:2004ps,Abazajian:2002ck}.  The meaning of
  the bracketed numbers is explained in the text.}
\end{table}

The limits on $g_{\epsilon\mu}^{\gamma }$ translate into order of
magnitude upper-limits
on the MPs.  Using the definitions in
Ref.~\cite{Hagiwara:fs}, and their limit on $(b+b^\prime)/A < 10^{-3}$
at 90\% CL as well as the fact that $A\cong 16$, we obtain the limits
given in Tab.~II.  The meaning of the numbers is explained in the
caption.  The bracketed limits on $\xi^\prime$ are not fully
constrained by upper-limits on neutrino mass.  They are included in
the table 
because the parameters with the largest uncertainties that enter its
definition are here better constrained.  In particular, the largest
uncertainty in
\begin{eqnarray}
1-\xi^\prime = [(a+a^\prime)+4(b+b^\prime)+6(c+c^\prime)]/A,
\end{eqnarray}
stems from the relatively large PDG upper-limits on the parameters
$a,a^\prime,c,c^\prime$ when compared to the upper-limit on
$(b+b^\prime)/A$.  Our limits on the former parameters is 
substantially better, thus improving on the PDG limit for
$1-\xi^\prime$ even though the neutrino mass upper-limit does not
constrain $(b+b^\prime)/A$.  With the improved limits on
$a,a^\prime,c,c^\prime$ due to Planck data, the upper-limit on
$1-\xi^\prime$ should then be entirely due to the upper-limit on
4($b$+$b^\prime$)/$A$.  In the same vein,  note that
because our constraint on $\alpha$ is so strong, the
measurement of
$\eta = (\alpha - 2\beta)/A$,
at PSI~\cite{PSI} to a few parts in $10^{-4}$
will also constitute a measurement of the MP $\beta$.

Finally, note that a similar analysis for the decay  $\tau
\to \mu ~\bar{\nu}_\mu \nu_\tau$ can be performed
\beq
\label{eq:tau-decay}
{\cal L}_{\tau \to \mu \bar{\nu}_\mu \nu_\tau}={4\gf\over
\sqrt{2}}\sum_{{\gamma=S,V,T\atop \mu,\tau=R,L}}
g_{\mu\tau}^{\gamma (\tau)}\bar
\mu_\mu\Gamma^\gamma\nu_\mu^n\bar\nu_\tau^m\Gamma_\gamma \tau_\tau~,
\eeq
the following limits are obtained:
$g^{\text{S}(\tau)}_{\text{RL}},g^{\text{V}(\tau)}_{\text{RL}},g^{\text{T}(\tau)}_{\text{RL}}<10^{-4}$ and 
$g^{\text{S}(\tau)}_{\text{LR}},g^{\text{V}(\tau)}_{\text{LR}},g^{\text{T}(\tau)}_{\text{LR}}<10^{-6}$.
In a particle physics model where the charged-lepton decay couplings
are all of the same order, the $g_{\mu\tau}^{\gamma (\tau)}$
should provide the best limits on the MPs.

{\bf Non-SM contributions to neutral currents.}
In light of the NuTeV result on the weak mixing angle
($\theta_{\text{W}}$)~\cite{Zeller:2001hh},
constraining non-SM neutral currents is particularly timely.  To determine
$\sin^2\theta_{\text{W}}$, the
experiment measures the ratio of neutral to charged currents in 
$\nu_\mu(\bar{\nu}_\mu)$-quark interactions.  Any deviation from the SM
neutral or charged current can be interpreted as a deviation
from the SM predictions for $\sin^2\theta_{\text{W}}$.  For neutral 
currents, the
relevant coupling constants are $a^{qq}_{i;ll^\prime}<
10^{-3}$ for $q=u,d$ and $i$=S,PS,T.  The (axial-)vector currents of the
SM do not change the chirality or the flavor of the neutrino while the
chirality-changing coupling
interactions considered in this work do.  Thus, the final states are
different and the rates---not the amplitudes---must be added.
Therefore, the chirality-changing non-SM operators can at most
modify the SM neutral current by $10^{-6}$ and can not account for
the NuTeV anomaly.

{\bf Conclusions.}
Derivation of our results requires only minimal assumptions. We view
the SM as an EFT valid below a certain energy scale (taken
to be above 1 
TeV) and assume the validity of perturbation theory.  Note that the interactions
of \Eq{eq:interact} are not gauge-invariant under SU(2$)_{\text{L}}\times \text{U}(1)_{\text{Y}}$.  From a strictly formal
point of view, our EFT is not allowed since $\mu$ is above the weak scale; 
the operators could be embedded in a gauge-invariant structure, but the resulting Ward identities may impose relationships between the parameters which are assumed independent in this letter.
However, since the neutrino mass does not violate gauge-invariance
(e.g., in the SM, the neutrino mass is generated through the spontaneous breaking of a gauge symmetry),
diagrams that contribute to $m_\nu$ are not forced to cancel in a gauge-invariant model.  
Our order-of-magnitude estimates should therefore be robust---finely tuned cancellations not withstanding.  The
MPs $\delta$ and $\rho$ will soon be constrained with  
improved precision by the TWIST experiment at the
$10^{-4}$ level~\cite{twist}.  
Although results of such measurements will be valuable whether or not a
positive signal is observed, an especially 
interesting situation would arise in the case where TWIST measured 
finite deviations from the SM values of
$\delta$ and $\rho$ since that would have implications for the neutrino 
mass.  Thus, any particle physics
model that could accommodate deviations of $\delta$ and
$\rho$ at the $10^{-3}-10^{-4}$ level would also be
challenged to simultaneously generate neutrino masses consistent
with observations; for example, this could be achieved through finely 
tuned cancellations of
the radiative corrections to the neutrino mass shown in
\Fig{fig:2loop} 
or mixing with right-handed neutrino states with masses $\gg$~0.23~eV that could lead to large contributions to the MP's.  
Furthermore, such a measurement would have 
implications for all physical processes where the magnitude of the
neutrino mass plays a role, like $\nbb$-decay when
the neutrino is a Majorana fermion.

\begin{acknowledgments}

The authors are grateful to Petr Vogel for very useful discussions and
observations.  The authors would also like to thank D. Bryman, 
S. Kettell, S. Pastor, R. Shrock, and A. Yu Smirnov for useful comments. 

\end{acknowledgments}

\end{document}